COMMUNICATIVE FIGURATIONS

Juliane Jarke, Gabriela Molina León, Irina Zakharova, Hendrik Heuer, Ulrike Gerhard

# Beyond Participation: A Review of Co-Creation in Computing

Universität Bremen | University of Bremen
ZeMKI, Zentrum für Medien-, Kommunikations- und Informationsforschung
Linzer Str. 4, 28359 Bremen, Germany, E-mail: zemki@uni-bremen.de
**www.uni-bremen.de/zemki/ | www.uni-bremen.de/en/zemki/**

**Juliane Jarke** (jarke@uni-bremen.de)

Juliane Jarke, PhD, is a senior researcher at the Institute for Information Management Bremen (ifib) and Centre for Media, Communication and Information Research (ZeMKI) at the University of Bremen.Beforehand she worked at the Centre for the Study of Technology and Organisation at Lancaster University.Her research focuses on public sector innovation, digital (in)equalities and participatory design.

**Gabriela Molina León** (molina.leon@uni-bremen.de)

Gabriela Molina Leon is a doctoral researcher at the University of Bremen. Her research focuses on Interactive Data Visualization and Computational Social Science. She studied Computer Engineering and Human-Computer Interaction at Universidad Simón Bolívar (Venezuela), INSA Lyon (France) und Bauhaus-Universität Weimar (Germany).

**Irina Zakharova** (izakharova@ifib.de)

Irina Zakharova is a doctoral candidate at the Centre for Media, Communication and Information Research (ZeMKI) at the University of Bremen and a research associate at the Institute for Information Management Bremen (ifib), Germany. Expanding on the notion of methods performativity, her doctoral thesis maps out the field of datafication scholarship with particular attention to research methodologies and concepts about datafication they produce.

**Hendrik Heuer** (hheuer@uni-bremen.de)

Hendrik Heuer, Dr., is a senior researcher at the Institute for Information Management Bremen (ifib) and Centre for Media, Communication and Information Research (ZeMKI) at the University of Bremen. His focus areas are Algorithmic News Curation, Disinformation, and Accessibility. He studied Human-Computer Interaction and Machine Learning in Bremen, Buffalo, Stockholm (KTH), Helsinki (Aalto) and Amsterdam (UvA).

**Ulrike Gerhard**

Ulrike Gerhard worked as a researcher at the Institute for Information Management Bremen (ifib) and Centre for Media, Communication and Information Research (ZeMKI) at the University of Bremen from 2016 to 2019. She has left academia and now works as a social worker.

# Beyond Participation: A Review of Co-Creation in Computing

## 1 Introduction

Over the past ten years, co-creation has become a buzzword for the development, design, and provision of products and services across a range of domains: in the private and public sector, but also in the areas of art and research. As such, "co-creation is widely understood as practices where a design practice and one or more communities of practice participate in creating new desired futures" (Lee et al., 2018, p. 15). Beyond this very broad understanding of co-creation as a future-making practice involving different "communities of practice" (Lave & Wenger, 1991) there is no one definition about what it is and how it ought to be done; the origins of the term and areas of application are manifold, and so are the objectives of projects relating to co-creation. It may be employed as a process, agenda or tool (Lee et al., 2018) involving different groups of people, for example, older adults (Jarke, 2021) or citizens more broadly (Nambisan & Nambisan, 2013b).

For the area of design, Sanders and Stappers describe co-creation as "the new landscape" and argue that "bringing co-creation into design practice will cause a number of changes to occur. It will change how we design, what we design, and who designs. It will also affect the tools and methods that the new teams of co-designers will use" (Sanders & Stappers, 2008, p. 15). This claim is equally true for other domains in which ideas about co-creation have impacted on traditional methods, roles, and objectives. For example, Voorberg et al. (2015, p. 1346) state that co-creation came to be considered "a cornerstone for social innovation" in the public sector. In the private sector, co-creation has become part of a business model that promotes customer-centricity through value co-creation (Piller et al., 2010).

In this paper, we address the problem of multiple, at times contradicting, definitions of co-creation across different domains relating to computing and in particular human- centred design. We do so by analysing different notions of co-creation present in the literature and by providing a framework for this analysis, attending in particular to the role of technology. So far, the role of technology has not been regarded sufficiently. However, we find that it makes a substantial difference on whether technology is understood as a tool to enable co-creation or as the objective and outcome of a co-creation process. More specifically, we address the following research questions (RQ):

**RQ1** In which contexts and how is the term co-creation used in computing and related disciplines?

**RQ2** What are the differences and commonalities between studies that self-identify as co-creation beyond the mere reference to participation?

**RQ3** What is the role of technology in different co-creation contexts?

We conducted a literature review of all papers in the ACM Digital Library—the world's largest scientific and educational computing society—whose authors attribute their work to co-creation and in doing so co-construct the term. The initial body consisted of 194 papers using the terms "co-creation" or "co-create" in the abstract. With these, we conducted a

keyword analysis to identify the most common topics. After an initial screening of the abstracts, we retained 62 papers for further analysis.

This paper contributes to our understanding of co-creation practices with respect to five dominant domains that we identified through our literature review: open data & public services, healthcare, learning, art and culture, and business. We introduce an analysis framework that allows to distinguish different understandings of co-creation, such as the role of technology or how the involvement of “others” is constructed and framed. Our framework implements broad coverage across domains and approaches when compared to the keywords used by the authors themselves. Based on the literature review, we identify three trends in co-creation work: Co-Creation as Co-Design Practice, Enabling Co-Creation through Digital Tools, and Co-Creation as an Ideal. We identify which of these types are most prevalent in the five domains and describe them to give an overview of the different perspectives on co-creation.

## 2 Related Work

Our work complements previous literature reviews on co-creation that have a focus on specific understandings or types of co-creation. For example, Lee et al. (2018) review ten design and innovation projects across a spectrum of application areas and propose a “design choices framework of co-creation”. Voorberg et al. (2015) review 122 reports covering co-creation and co-production in the public sector. In a survey of 493 papers, Degnegaard (2014) identifies five streams of co-creation comprising of shared meaning, shared values, technological platforms for further collaborative work, new products or services and design, relating to five different areas of application. Overall, the analytical framings of such reviews differ with respect to examining types of design decisions (Lee et al., 2018); shifting roles and changing methods (Sanders & Stappers, 2008); and objectives and outcomes (Voorberg et al., 2015), each in specific fields. By contrast, we focus on analysing the role of technology in different co-creation contexts, and characterise the application domains accordingly.

In the following, we provide a brief overview of the main concepts related to co-creation: (1) in the private sector, the co-creation of value defines a new, customer-centric business model; (2) in the public sector, co-production or co-creation of public services allows for social innovation and citizen-centric service development and delivery; (3) across different sectors, co-creation—as an extension of co-design or participatory design—transforms computing practice. In all of these approaches, the participation of external stakeholders plays a key role, for example, customers, citizens, and users. Their relative importance is reflected later in the findings of the keyword analysis (see Results section).

*Co-creation of Value*

In management and business literature, co-creation is described as a business model that has become relevant in the context of changing markets. Due to individualisation processes, consumer profiles have differentiated and multiple new ways of obtaining feedback from consumers are required. These changes are expressed through ideas such as “customer centricity” (Piller et al., 2010, p. 4) and “service-dominant logic” (Roser et al., 2009, p. 6). Instead of creating or influencing a demand, the “customer-centric firm” (Piller et al., 2010, p. 7) responds to existing demands and uses external knowledge to develop innovative solutions.

While co-creation can be seen as “any form of customer involvement” (Roser et al., 2009, p. 7), it has been recently discussed as a new mode of interaction between suppliers and customers. More specifically, Prahalad and Ramaswamy (2004) define value co-creation as the process in which customers create value with companies proactively. The value should benefit both the customers and the companies, and the process occurs when they engage in collaborative service design. This distinguishes co-creation from other forms of involvement, such as competitive "crowdsourcing" for a solution to a specific problem (Roser et al., 2009, p. 11), or the "lead-user-concept" (Piller et al., 2010, p. 9) where companies look for existing solutions developed by customers/users. Co-creation is defined as a longer process and an intense and equal relationship initiated and facilitated by an organisation (Prahalad & Ramaswamy, 2004, pp. 9–10). Different modes of co-creation processes can be described, in regard to the following aspects: (1) Involved parties: Phases/stages in the development process (from needs to solutions); (2) Degree of involvement/collaboration (structure of relationship, frequency of interaction); (3) Degree of freedom (purpose, openness/predefinition of tasks, predictability of the solution). This list illustrates how the business and management literature largely disregards technology and its role in processes that aim to co-create value.

*Co-design and Participatory Design*

In design literature, the term "co-design" refers to a specification of co-creation that ascribes "collective creativity across the whole span of a design process" with a focus on the collaboration between designers and (future) users (Sanders & Stappers, 2008, p. 6). Co-design differs from traditional approaches to user-centricity as it extends its intention, and instead of designing *for* users, it aims to design *with* users for "future experiences" that consider actual needs of individuals, communities, or societies. In this context, the focus shifts from the *designing of* "products" to *designing for* a "purpose" (Sanders & Stappers, 2008, p. 11).

There are at least three different origins of co-design or participatory system design: The UK-based approach and prominent ETHICS-Method (Mumford, 1981; Mumford & Henshall, 1979), the Scandinavian approach with the well-known DEMOS and UTOPIA projects (Ehn, 1988), and the US-American approach on "Cooperative Design" (Greenbaum & Kyng, 1991). In this tradition, the focus on the involvement of future users and their "collective creativity" (Sanders & Stappers, 2008) is idealistically applied throughout the entire design process and changes the roles of users from design subjects to partners (Sanders & Stappers, 2008). Druin (2002) describes the level of stakeholders' involvement along a four-dimensional model, including new roles such as users, testers, informants, and design partners.

The goals of participatory design include moral as well as pragmatic considerations. Vines et al. (2013) list the following aspects which—to a differing degree—build the rationale for participatory design projects: (i) the sharing of control with users; (ii) the sharing of expertise and (iii) individual, organizational, and technological change.

*Sharing control with users.* Rooted in the political agenda of Scandinavian participatory design, one of the main goals of participatory approaches is the destabilisation of power structures by sharing control over the design process and outcome (Vines et al., 2013). Humans ought to be regarded as "actors", not "factors" (Bødker, 2006). It is hence important to consider the institutional framing of participatory projects in order to understand "the sources of power and influence different project participants were able to mobilize" (Bratteteig & Wagner, 2016, p. 429). This includes considerations about the (hidden) agendas participants may have.

*Sharing expertise.* Including future users' input in the design process also makes pragmatic sense as it is said to increase the chances of a successful design outcome by taking into account their "expert perspectives and preferences regarding the activity that the design will support, and most likely transform" (Carroll & Rosson, 2007, p. 243). One of the most common ways of eliciting users' expertise are workshops in which teams of researchers, designers, system developers, future users, and other stakeholders come together to identify challenges and develop new ideas. In these workshops "boundary objects" (Star & Griesemer, 1989) are co-developed that act as "shared articulations of knowledge of those participating in the design process" (Vines et al., 2013, p. 430).

*Enabling individual, organisational, and technological change*. Finally, all participatory design approaches recognise that such processes are motivated by enabling (or enforcing) some kind of change. Participatory design hence needs to understand peoples' current practices, experiences and how future design products may be appropriated (Vines et al., 2013).

The origins of the concepts such as co-design and participatory design can be seen as a reason why the main focus of co-creation in this line of research is on the involvement of various stakeholder groups, power dynamics as well as on societal change through co-creation. In this context, technology may represent both 1) a powerful *(enabling) tool* to overcome existing power structures in a co-creative process and 2) an objective (*output*) of creative collaboration with potential for change. From this point of view, attention needs to be place on the role of technology in such participatory and co-design projects.

*Co-production*

The co-production of services has received an increasing amount of attention by public sector stakeholders against the backdrop of financial cuts, the increasing complexity of societal challenges, and the availability of new technologies (European Parliament, 2014). Ansell and Gash (2008) have called this approach "collaborative governance". Following collaborative governance, public agencies directly engage non-state stakeholders in collective decision-making processes. As Aichholzer and Strauss (2015) note, this definition includes all stages of the policy lifecycle and includes citizens as individuals as well as organised groups. The involvement of citizens as service users and their communities in the design and delivery of public services is also often referred to as co-production (from service planning, design to delivery). Furthermore, co-production is employed to foster the inclusion of disadvantaged or marginalised groups of citizens and refers to the collaboration of at least two stakeholder groups, public administrations, and citizens. Co-production has to be distinguished from other non-governmental forms of civic self-empowerment such as volunteer work or self- organisation.

Co-creation in the public sector as an impact-oriented form of collaboration between public administrations and citizens aims to unfold the capacities, potentials and strengths of all parties concerned with the objective of enhancing life quality in neighbourhoods, cities or regions, and to jointly achieve efficiency gains (Löffler, 2015, p. 319). Focusing more on the relationships engaged by co-production activities, another definition refers to co-production as 'the provision of services through regular, long-term relationships between professionalised service providers (in any sector) and service users or other members of the community, where all parties make substantial resource contributions' (Bovaird, 2007, p. 847). Hence, co-production of public services refers to the long-term involvement of citizens in problem definition and solving. In their analysis of public sector co-creation examples and case-studies, Nambisan and Nambisan identify four different roles in the practice of co-creation for citizens (Nambisan & Nambisan, 2013a):

- Explorer (identifying, discovering, or defining a problem)
- Ideator (conceptualising solutions)

- Designer (designing and developing implementable solutions)
- Diffuser (implementing the solution)

Voorberg et al. (2015) distinguish three different roles citizens may assume in co-producing along the life-cycle of service planning, design and provision: citizens as initiators, co-designers, and co-implementers.

According to Voorberg et al.’s review of 122 reports, among the top mentioned goals of co-creation, authors list more effective (18%) and more efficient (11%) services. Of those reports that report the outcome of their projects, 25% mention an increase in citizen involvement. However, half of the reports (52%) do not mention an objective at all. Voorberg et al. embarked on their review in order to identify whether the big hopes for co-creation — they speak of a “magic term"— can be based on evidence in order help public sector decision makers, decide whether and how to initiate such processes. In sight of this review, Voorberg et al. argue that it is not clear whether co-creation does indeed contribute to the outcomes it aims to deliver. They further question, “if there is a relationship between several degrees of citizen involvement (co-implementing, co-design and initiator) and the outcomes of social innovations” (p.1348). In conclusion, they argue that in the majority of cases, co-creation seems to be considered as a virtue in itself.

Overall, there is a dominance of co-produced public services in the areas of health care and education with most services not involving digital components. In the review by Voorberg et al., the dominant factors that manage to achieve a successful sense of co-creation are either organisational (e.g. civil servants’ attitude towards citizen participation or a risk-averse administrative culture) or they can be found in the circumstances and in the actions of participating citizens (e.g. their socio-demographics, motivations, and social capital). It is subsequently noticeable that the role of technology is not considered well in this line of work and that there is a need to investigate it as a potential enabler for public service co-creation and/or as the outcome of a co-creation process in form of a digital public service.

*Summary*

All of the approaches discussed above consider the involvement of stakeholders outside a traditional frame (e.g. users in co-design, citizens in co-production, customers in value co-creation) as beneficial for co-creating joint futures (e.g. future services or products). Those collaborative activities are entangled in socio-technical assemblages such as digital service provision. However, the role that technology may play is not well explored. Technology may *enable* the co-creation of new products and services; new technological design ideas or products may also be an *output/result* of co-creation. In the first case, then, technology can be seen as *means* for co-creation and in the second case as an *objective*.

To review the meaning of co-creation and to help us understand how co-creation is practiced, we focus in our review on both the role of co-creation’s target audience *and* the role played by technology in collaborative future making. In the following section we describe our research methodology and the dataset used to answer those questions.

## 3 Methodology

We conducted our survey in the ACM Digital Library, the world’s largest scientific and educational computing society. We only reviewed papers that used the terms co-creation or co-create in their abstract. This means that papers about co-creation that do not refer to the term in their abstracts are missing. Our focus on the specific term also excludes work that strongly relates to co-creation without explicitly referring to the term. The reason for this is grounded in our social constructivist approach: For this study, we are interested in how authors—who self-identify their work as co-creation—describe the process, objectives, and outcomes of co-creation and construct the term by providing empirical examples and evidence on their co-creation practices. We do not set out to provide a clear definition of what co-creation is, but, rather, aim to understand how the term co-creation is used and why.

Our search in the ACM Digital Library yielded 194 results, from 2000 to 2018. The first step of our inquiry , involved a keyword analysis to identify the main research clusters. All keywords were retrieved from the ACM and assigned by the authors of the papers. Overall, we found 680 unique keywords. For our investigation, we merged similar keywords such as "codesign" and "co-design", "probe":"probes", "vr":"virtual reality", and "wearable technologies" and "wearable computing". This task was carried out by two authors of this paper who reviewed all keywords and marked those that should be merged. In total, we merged 33 keywords. For the purpose of our analysis, we focus on the 36 keywords that occur three times or more. This means that the vast majority of the keywords, more than 646, only occur once or twice. Regarding the wide distribution of keywords, we conducted a further in-depth analysis of whole paper bodies.

Narrowing down the search, all 194 abstracts were read twice by two different authors. We retained only papers and excluded all other publication types such as keynotes and workshops. From the paper abstracts, we excluded all those which used co-creation in a colloquial way (e.g. Jacq et al., who describe how humans and robots ‘co-create’ the content of a short story (Jacq et al., 2018)). Of the initial 194, 90 papers were retained.

The next step was to define those papers relevant for our comparative study. We read all remaining papers and assigned them as relevant if they matched all of the following characteristics. First, sufficient information on the collaborative engagement of different stakeholders should be provided as a prerequisite characteristic of co-creation (e.g. descriptions of the stakeholders and of their involvement in co-creation). Secondly, co-creative activities should pose the focus of the project or research rationale. All selected papers were coded alongside categories such as the role of technology as a means or objective of co-creation, tools and methods to support co-creation, target audience and the types of technology implemented or developed in the co-creation process. We give a more detailed description of our analysis framework in the following section. Alongside the coding process, we also took note of each paper relevant to its domain. These domains were then clustered (e.g. care services, public health, home-based therapy were assigned to the cluster health care).

Preliminary results showed that much of the co-creation work is domain-specific. For this reason we only analysed those domains with more than five papers specific to a particular domain. Domains with less than five publications include ICT4D and postcolonial computing, urban design and architecture, formal education, sustainability and the environment, prisons, and sports. For these domains, no area-specific characteristics of co-creation could be identified. We did not merge the learning and formal education domains as the latter has a stronger focus on educational institutions and infrastructures whereas the first concerns learning activities and the experiences people acting individually.

This left us with 62 papers from five application domains: (1) open data & public services, (2) health care & ambient assisted living, (3) learning, (4) art & culture, and (5) business. In order to validate our selection, we used the keywords of the initial 194 papers to determine how much coverage we have once the filtering steps were completed.

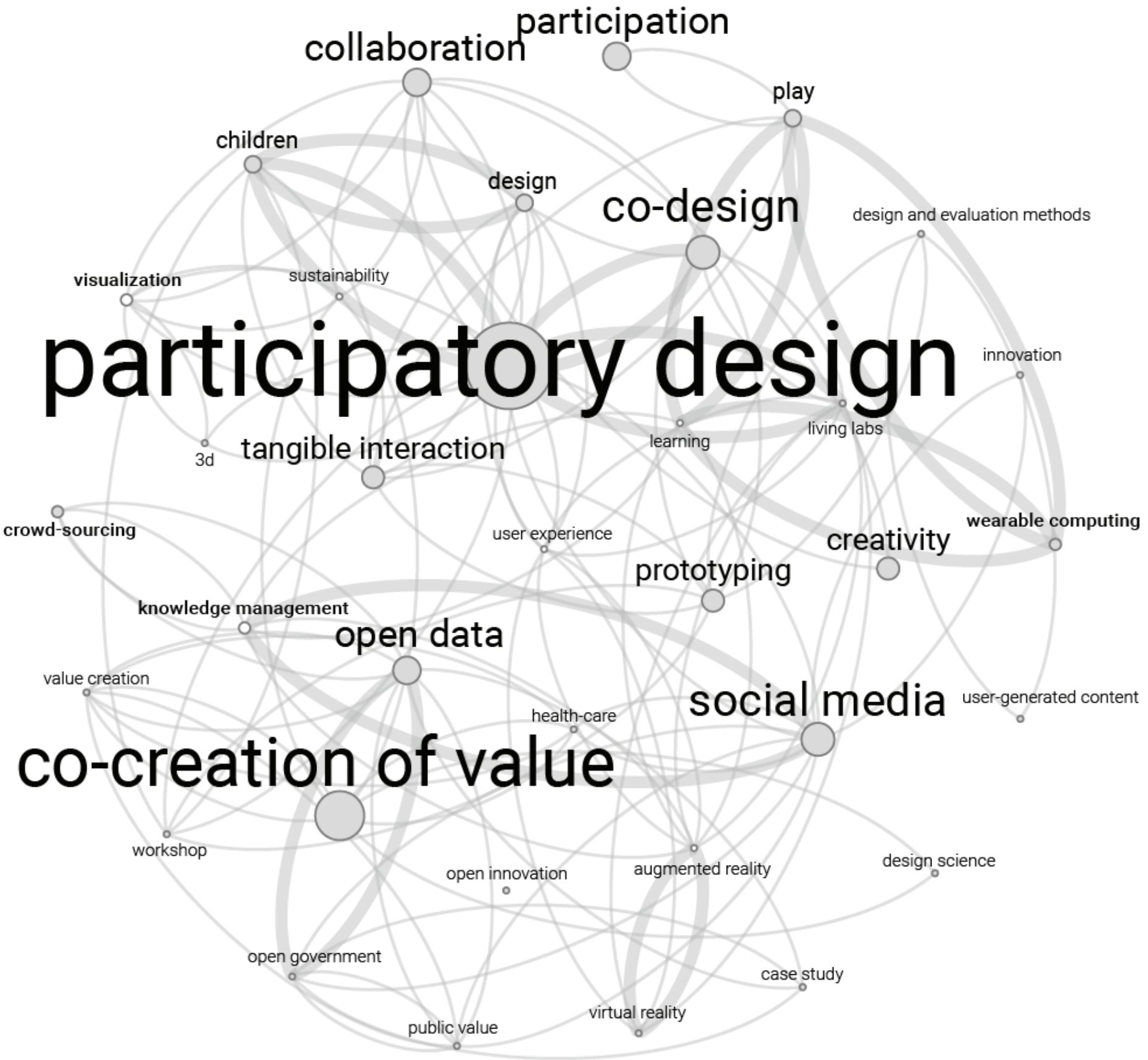

*Figure 1: The graph visualizes the author keywords of all 194 papers that use the term co-creation in their abstracts. The size of the words is proportional to how often they occurred. The edges between keywords indicate how often words co-occur. The nodes of keywords of relevant papers are coloured grey and the others white. 'Co-creation' as a keyword has been excluded.*

## 4 Results

*Counts of keywords*. Figure 1 shows a graph of the 36 keywords that occur three times or more in the initial body of 194 papers referring to co-creation in the abstract. Each node in the graph represents a keyword, and each edge indicates a co-occurrence of two keywords. The graph was generated and visualized using the network analysis software *Gephi*. The size of each node is proportional to the times the keyword was used across the 194 papers. The labels of the two least frequently appearing nodes (3 and 4 times) was increased to improve readability. For the graph, we removed the most frequent keyword "co-creation", which appeared 30 times. The most frequently assigned keywords other than co-creation are participatory design (18), co-creation of value (11), social media (8), and co-design (8). The keywords collaboration (7), open data (7), participation (7), creativity (6), tangible interaction (6), and prototyping (6) are also frequently assigned. To illustrate the differences between the initial body of 194 papers and our final selection of 62 papers, we marked the nodes of the keywords used in the 62 retained papers in grey. The ten most frequently used keywords each appear in one or more of the retained papers. This means that despite the filtering steps, we have a broad coverage over the body of co-creation literature in the ACM Digital Library. These keywords demonstrate that most papers trace the origins of their use of co-creation to either participatory design and co-design or the co-creation of (public) value. The keyword analysis demonstrates that there is a clear reference to technology either as an *enabler* for co-creation (e.g., social media) or as its *objective* (e.g., tangible interaction). This further supports our claim there are analytical benefits to be had when considering technology’s role in co-creation. In addition, papers cite *co-creation methods* (e.g. prototyping, crowdsourcing, workshop, living labs), *paradigms* (e.g. participation, creativity, learning), and *specific target audiences* (e.g. older adults, children).

Based on the author keywords, we also analysed which words co-occur with which. The most important finding is that most keywords only co-occur once with another keyword (94.4%). Only 5.1% of the terms co-occur at least twice and only four words co-occur more than twice. All of them co-occur either with the term co-creation or co-design. We found three co-occurrences for the pair of co-creation and creativity, as well as for co-design and older adults. The latter co-occurrence suggests a particular interest in co-designing for aging populations. The terms participatory design and open data co-occur with co-creation four times. This suggests that although these papers relate to open (government) data, their focus is not the co-production of public services, but, rather, participatory design based on open data.

*Keywords of irrelevant papers*. Six of the 36 keywords that appear three times or more are not assigned to any of the relevant papers. These keywords are: knowledge management (4), visualization (4), sustainability (3), augmented reality (3), user experience (3), and learning (3).

*Framework*

The following framework describes the categories we used for the coding of relevant papers. It stems from the review of related work and its different understandings of co-creation. This relates to the framing of what stages a co-creation process may cover (from planning and problem definition to co-creation of the anticipated output, a service, technology or value, for example); the role of technology as enabler and/or objective of co-creation and the definition of the co-creation target audience and appropriate methods. The results from our keyword analysis support this selection of key co-creation dimensions as does it contribute to the conceptualisation of co-creation approaches in the different fields of research.

*Conducting co-creation*. Through our initial screening of abstracts, we realised that many papers did not report co-creation activities and did not conduct co-creation themselves. It seems important, therefore, to distinguish between contributions implementing co-creation methods and those who do not so as to understand why the latter refer to co-creation as part of their research.

*The role of technology*. The appearance of terms such as social media and augmented reality in the keywords analysis demonstrate the presence of specific technologies in at least a subset of the papers. The related work suggests that the role of technology differs across approaches to co-creation. While technology is often the goal of co-design and participatory projects, it may have a different purpose in contexts concerned with value co-creation or co-production. For this reason, we analysed the role technology plays across each different domain, distinguishing between technology as a means for co-creation or its objective.

*Enabling co-creation*. If technology is conceptualised as a means for co-creation, the focus shifts from the emergence of a technology to its use (the functionality and implementation of certain tools are described, for example). Apart from technology, the keywords suggest that there are multiple participatory design or co-design methods that facilitate or enable co-creation (such as prototyping, crowdsourcing, living labs).

*Co-creation output*. Stemming from our review on related work and the diversity of domains, it is to be expected that the target audiences and the results of co-creation processes differ. Tracing those categories while reviewing co-creation may help to detect possible patterns across the different domains and approaches to the role played by technology.

## 4 Analysis

*Overview*

We analysed all 62 papers based on the framework provided above, and the insights we gained by analysing the Sankey diagram in Figure 2. As can be seen in the diagram, we have clustered and color-coded the papers according to the five dominant domains (first category on the left-hand side). The second column visualizes whether the papers have conducted their own co-creation activities (and report on these) or not. The third column visualizes the role of technology; whether it is understood as an objective of co-creation or as a means to co-create. The cases where technology was considered as both or did not appear at all are also represented. In the fourth column, we visualize what papers have described as enablers of (their) co-creation activities: specific co-creation methods, digital tools, both or none. The fifth column provides an overview of the co-creation outputs. Finally, we indicate the target audiences. For further exploration, an interactive version of this diagram is available under: https://cocreation.uni-bremen.de/fig2.html. Within what follows is an analysis on each individual domain before we summarise and discuss our findings in regard to co-creation across domains.

*Open Data & Public Services (n=14)*

Papers in this domain are concerned with the (re)-use and co-creation of open (government) data as well as the co-production of public services. There are a number of papers that describe the use of open data platforms for the co-creation of public services or for the co-creation of open data.

In most papers, technology is understood as a means of co-creation. The technologies are in most cases platforms (e.g. open data platforms, social networking platforms) and to a lesser degree, mobile applications. In those papers attending to the co-creation of public services without specific reference to open data, technology is being used as a means to co-create knowledge (see, Giesbrecht et al., 2015) or to foster citizen participation (see, Omar et al., 2018 on participatory budgeting). Paper like these describe, for example, the use of digital platforms (such as social media) to co-produce or co-deliver services either for citizens or for the government (Chouikh et al., 2016).

Those papers focusing on open (government) data, stress the importance of co-creation of public value, relating either to the co-creation of local knowledge or digital public services relevant to specific groups of citizens (e.g. Cordasco et al., 2017; Callinan et al., 2018). Research questions of these papers concern the ways in which citizens can be stimulated to use and/or co-create open government data.

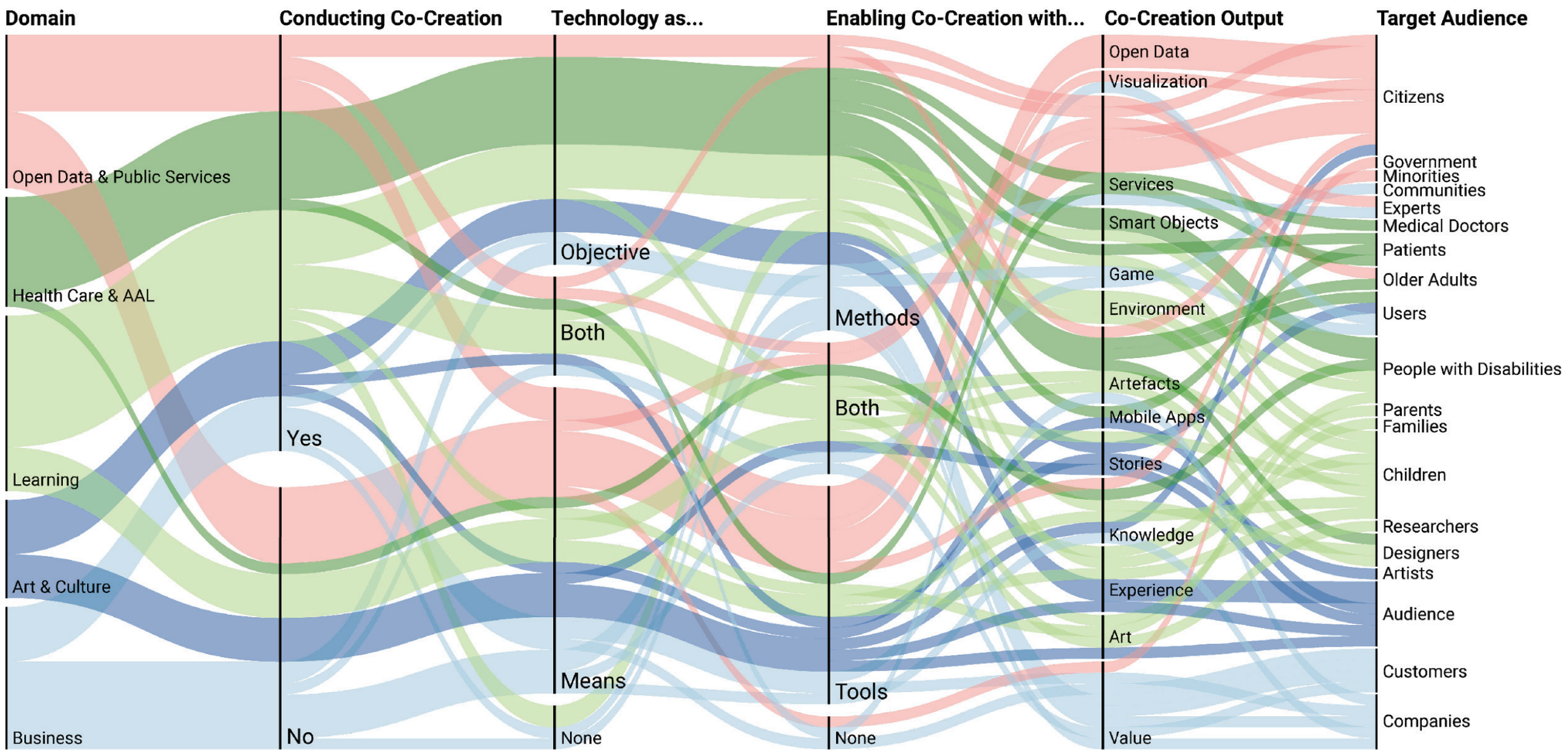

*Figure 2: Sankey diagram that visualizes the distribution of the 62 relevant papers through the four framework categories. To give a complete overview, we add the recognized domains in the first column – encoded using different colours, and the target audience of each of the co-creation domains in the final column. For an interactive version of the figure see: https://cocreation.uni-bremen.de/fig2.html*

The ways in which this may be achieved are manifold, for example, some papers propose technological solutions such as platforms (Cordasco et al., 2017; Ambrosino et al., 2018) whereas others propose methods that use open data for the co-creation of dedicated public services (Scherer et al., 2015; Bull et al., 2017; McBride et al., 2018).

Only about half of the authors in our sample have conducted co-creation projects themselves. Three are conceptual or review articles on co-creation in public service design and delivery. The majority of papers target citizens as their main audience. Some also list government and otherpublic administrations as well as businesses. In those papers in which co-creation was carried out, it usually covered smaller groups of citizens engaged either in workshops and through interviews or via collaborative platforms.

The role that co-creators assume differs between open data papers and those related to public services in general. For the open data papers, citizens as co-creators are mainly involved as data creators (also sometimes referred to as data crowdsourcers). In those cases, the co-creative work is facilitated through a platform. In the cases where technology is the goal of co-creation, methods such as workshops and interviews are the most common.

Some of the papers have been published early on during the projects, so, rather than analysing co-creation projects, they report on pilot studies and planned co-creation activities.

*Health Care & Ambient Assisted Living (n=10)*

Papers in this area are concerned with the co-creation of supportive technologies for people with disabilities (AAL) or technologies that facilitate personal and public health care. In most cases, co-creation is described as design practice for the development of technologies that fit the specific needs of a particular target group. However, we also found cases where co-creation is understood as an ideal for collaborative health care.

In the area of AAL, technology is the objective of co-creation processes. Assistive technologies are co-developed in order to support everyday activities (in particular mobility) of people with disabilities (in particular blind people) (Branham & Kane, 2015; W. C. Ugulino & Fuks, 2015; W. Ugulino & Fuks, 2015). The same applies for co-creation processes aiming at the development of technologies that facilitate personal health care. Co-creating for personal health care aims at either supporting the personal treatment of patients with a particular disease (Duval et al., 2018; Stütz et al., 2017), the reporting and/or visualization of patient data (Ivatury et al., 2017; Sauvé et al., 2017) or to support active and healthy aging (Foverskov & Binder, 2011). In all these cases, co-creation is understood as a collaborative design practice where the technological outcome facilitates more independent forms of living, therapy, or preventive health care of a (more or less) specific target audience. However, personal health care is also promoted in the public sector (Moll, 2010). Here the participatory development of technologies is seen as a means to enable co-creation in a much broader sense: A role shift between citizens and/or patients from mere service users to co-creators of health services is, therefore, aspired to.

*Learning (n=16)*

In the educational domain, papers address technology as means of co-creation and technology as the objective of co-creation almost equally often; three papers deploy technology as both (Heljakka & Ihamäki, 2017; Rubegni & Landoni, 2018; Yuill et al., 2013). Types of technology vary between tangibles and web applications/platforms (respectively five and four papers) and the Internet of Things, mixed reality as well as mobile applications. Although three of four papers referring to web-based applications and platforms understand technology as a means of co-creation, there is no clear relation between the type of technology and its role in the co-creation process. The area of Children-Computer Interaction (CCI) stands out from other areas in this domain because of the high proportion of studies involving children in co-creation (12 from a total of 16). Nine of those papers aim at collaboratively developing tools and methods for enabling co-creation with children in general, or with a specific focus on children with disabilities (Brown et al., 2016; Frauenberger et al., 2016) or from disadvantaged families (Itenge-Wheeler et al., 2016). In the co-creation process, children are the main target group and they often take part in user tests, whereas their teachers, guardians or design teams take the role of ideators/design experts.

Children also take an active part in prototyping although only a few studies bring the co-created products to life and provide them for actual use (Frauenberger et al., 2016). Remarkably only few research teams involve the same children over a longer period and foster a relationship in which children become actual design partners. An exception is a study by Frauenberger, Makhaeva and Spiel (Frauenberger et al., 2016) in which autistic children become design partners as they not only conceive initial ideas for their personal smart objects, they also produce prototypes. Target audience involvement at the very early stage of the development process is, therefore, an option but not a must-have for co-creation in CCI. Only four of the 16 papers target whole families as future users and mostly aim at collaborative artistic experiences and gaining new knowledge (Herstad & Holone, 2012; Yuill et al., 2013; Krishnaswamy et al., 2017; Banerjee et al., 2018). At the same time, the families are actively involved in later stages of co-creation processes (such as user testing).

Considering our investigation of the keywords, it is surprising that the keyword "learning" was only used by papers we deemed irrelevant. One reason may be that the term

"learning" alone can be considered too generic. More specific terms, such as "experiential learning" and "learning artefacts" appear in the retained papers, however. For example, one paper that used the keyword, developed a wearable interface for music (Wilde, 2012).

*Art & Culture (n=9)*

Nine papers have a focus on co-creation in the domain of art and cultural heritage, while seven of them understand technology as means for designing collaborative artistic experiences or art (Bongers & Mery, 2011; Fuks et al., 2012; Han et al., 2014; Petrelli et al., 2016; Song & Jun, 2017; Taylor et al., 2011; Watkins & Russo, 2005). Only one paper aims at designing social media for a museum to engage visitors in new cultural experiences (Watkins, 2007). Interestingly visitors do not partake in the co-creation process in that study, but museum employees act as experts for idea creation and design. By contrast, most other studies involve visitors in co-creation by embedding design activities in exhibitions. At the same time, only half of the papers conduct co-creation activities, whereas others develop tools to enable further collaborative cultural experiences. The understanding of co-creation in art & culture differs from other domains at the level of stakeholder involvement in the later stages of co-creation, mostly involving them in producing collaborative experiences, but not technological solutions. Three studies also conducted and proposed co-creation methods specifically for museums (Fuks et al., 2012; Petrelli et al., 2016; Watkins & Russo, 2005).

*Business (n=13)*

There are thirteen papers related to the business domain, which represent 21% of the literature. Most of the papers refer to co-creation as a method for service design (Athavankar et al., 2014) and define services as processes of value co-creation (Lessard & Yu, 2012). In these papers the co-creation of value is understood as the interaction between service providers and customers, where listening is key to gathering the ideas and feedback of the stakeholders (Briggs & Makice, 2008), ideally leading to mutual benefit (Ojuri et al., 2018).

Service science plays an important role, as it is said to be "centered on the study of value co-creation within and among service systems” (Ojuri et al., 2018). Furthermore, Briggs and Makice (2008) introduce the term “deep co-creation" to refer not only to the co-creation of products and experiences by an organisation, but also to the co-creation of the organisation itself.

Online platforms are often used as a virtual space for co-creation (Briggs & Makice, 2008; Cu et al., 2016) where users can interact, generate, and share content. Eight papers fall into this category, in which technology is used as a means for co-creation. For example, Kärkkäinen et al. (2011) analyse the role played by social media on knowledge co-creation, while Goel and Mousavidin (Goel & Mousavidin, 2007) describe how companies can make use of 3D virtual worlds to approach new customers and co-create products and ideas digitally, before releasing them into the real world.

In eight papers, the objective is to co-create value in the context of service design and development. Kärkkäinen et al. (2011) study knowledge co-creation with customers on social media as part of an innovation process. Also, in the context of innovation, Huhtamäki et al. (2013) co-created visualizations to demonstrate the structure and dynamics of an open innovation platform for presentation and marketing. Two papers refer to the co-creation of game design where the players are seen not only as users or customers, but also as a community that co-designs the game. In Poretski and Arazy (2017), the user community co-created modifications to a game that later contributed to increased sales. Finally,

Briggs and Makice (2008) set the design of digital artefacts as a co-creation goal in the context of "deep co-creation".

In only five papers was co-creation conducted as a participatory design approach. Huhtamäki et al. (2013) applied it to design network visualizations and animations in collaboration with the representatives of an open innovation platform, to describe and promote it. Others conducted expert interviews (Tung et al., 2009) and focused on the development of artefacts for services (Tung et al., 2009; Briggs & Makice, 2008).

The main target audience were companies and their customers. Value co-creation was part of the relationship building process between businesses and customers. Goel and Mousavidin (2007) describe co-creation with customers as an approach in which customers are expected to proactively participate in the exercise of problem solving. In the case of games, the players are considered co-creators who interact virtually, providing feedback and suggestions to improve gameplay (Samper-Martinez et al., 2015).

*Summary*

Overall, our analysis demonstrates that there are not only multiple and differing understandings of co-creation, but also a manifold of different outputs and desired outcomes. In general, technology is embedded in every domain as an enabler of co-creation, as its objective or both. Web- or mobile platforms are more likely to be used to enable future co-creation whereas tangibles, wearables, or Internet of Things solutions are more often the output and defined as an objective.

The co-creation approaches differ further with respect to how specific they define the target audience of their co-creation outputs and the co-creators involved. In those domains, in which co-creation is primarily aiming at co-creating technology, the target audience is defined rather specifically (e.g. disabled people, children, families) attending to their specific needs. Within these kinds of co-creation projects, co-creator's knowledge and expertise about their everyday life or profession, among many other facets of their lives, is particularly important. By contrast, in those domains in which technology is understood as an enabler of co-creation (such as social media platforms that enable knowledge or value co-creation), the target audience is defined in very broad terms (customer, business, citizens, users).

Few papers in the business domain have conducted and described their own co-creation activities. While in all other domains, we find papers that are either conceptual or present reviews and have not conducted co-creation projects themselves, most papers in each of the domains (with the exception of business) present some empirical insights. This is particularly striking as many papers in the business domain propose methods for co-creation (some present digital tools in addition as well).

## 5 Discussion

All papers describe co-creation as engaging "others". These "others" were traditionally not involved in the creation of outputs and outcomes such as technologies, services, or knowledge. There is, then, an implicit understanding of whose participation is traditionally considered legitimate and part of a particular community of practice (e.g. designers) and whose participation is peripheral. By involving "others", the analysed co-creation approaches explicitly value the knowledge and expertise of all stakeholders (e.g. users in the case of co-creating technology, citizens in the case of co-creating public services or customers in the case of value co-creation). How such an involvement or engagement may operate, differs across the domains. For example, in most cases co-creators are meant to be involved in the definition (co-creation) of a problem, and not only in the discussion about a possible solution. Notable exception to this are the papers in the domain of art and culture, where (digital) experiences and stories are co-created.

The large majority of the papers in our investigation explicitly or implicitly assume that co-creation is of value. While some papers discussed challenges connected to co-creation, none of the papers criticised or opposed the concept. A cautious interpretation of this is that co-creation is a powerful imaginary that a large majority of actors agree is valuable. A bolder interpretation could regard co-creation as an almost moral imperative that stakeholders should be deeply involved and that this involvement inevitably creates value for all stakeholders. Such underlying assumptions about the intrinsic value of co-creation connect to the pragmatics and politics of research ethics and funding. For example, most of the papers around open data and co-creation are based on EU-funded projects aiming to promote work in the area of citizen engagement and co-creation around open government data (MobileAge - (Bull et al., 2017); ROUTE-TO-PA - (Ambrosino et al., 2018; Cordasco et al., 2017; Ojo et al., 2018); OpenGovIntelligence - (McBride et al., 2018; Stasiewicz et al., 2018); EMPATIA - (Omar et al., 2018)). This means that 7 out of the 14 papers we have reviewed in this domain have indicated that they have received EU-funding. This raises questions concerning how, through particular funding mechanisms, the European Commission (or other funding bodies), promote the research and innovation agenda of this particular topic area.

Overall, the investigation of keywords demonstrated that the most frequently used keywords are fairly generic, for example, collaboration (7), open data (7), and participation (7). Based on our analysis, we reviewed all papers and how they fit inside the framework. This allowed us to identify three salient types of co-creation, which helped to better interpret the keyword analysis (for a summary see table 1):

*Co-Creation as Co-Design Practice (Type A):* Papers of this type conducted and reported on their co-creation activities. The objective of these kinds of co-creation approaches is the co-design of technology. Papers in this category describe methods for co-creating technology. This type of co-creation is particularly prevalent in the domains of health care & AAL as well as learning. They define their target audience rather specifically (e.g. people with disabilities, medical doctors, families) as they situate technology design and use in the context of everyday life and user experience.

| | **Co-Creation a Co-Design Practice (Type A)** | **Enabling Co-Creation through Digital Tools (Type B)** | **Co-Creation as Ideal (Type C)** |
|---|---|---|---|
| Goal of co-creation activities | Co-design of technology, responding to the context of participants’ everyday life/work | Co-creation of services, (open) data or art/experience | Co-creation of value and/or knowledge |
| Role of technology | Technology as objective of co-creation | Technology as means to enable co-creation | Technology as a means to enable co-creation |
| Main domain(s) | Health care and AAL, learning | Open data and public services, arts & culture | Business |
| Definition of target audience | Specific target audience (e.g., people with disabilities, medical doctors, families) | Stakeholder groups (e.g., government, citizens) | Broad definition (e.g., users of a specific platform) |

*Table 1. Main differences and commonalities, found in our literature review (n=62), between studies that self-identify their work as co-creation across different domains relating to human-centred design.*

*Enabling Co-Creation through Digital Tools (Type B):* Papers of this type understand technology as a means for enabling co-creation. They describe tools/technologies that facilitate the co-creation of services or open data. This type of co-creation is particularly prevalent in the domain of open data & public services.

*Co-Creation as Ideal (Type C):* Papers of this type did not report on conducting co-creation activities themselves (as in type A), but, rather, describe methods or technologies that may facilitate future co-creation processes. They define their target audiences in the broadest terms possible (e.g. users of a social media platform). This type of co-creation is particularly prevalent in a business context aiming to co-create value or knowledge.

These three types broadly match the areas discussed in the related work section with Type A matching Co-Design, Type B matching Co-Production and Type C matching Co-Creation of Value. They differ however with respect to the specific attention to technology (not as important in the domain of co-production or co-creation of value). In terms of future research, we propose to investigate further whether the difference in the definition of target audiences also plays out with respect to the roles they assume (e.g. as explorers, ideators, designers, data curators, and/or users).

## 6 Conclusion

In this paper, we have provided a review of 62 papers that focus on co-creation. We have analysed and discussed similarities and differences of co-creation approaches across five domains. These were subsequently analysed along a framework to identify the role played by technology (means/objective), whether co-creation is conducted, how co-creation is meant to be facilitated, and what the expected output and outcome of co-creation may be. This will allow future co-creation projects to reflect and critically analyse their own understanding of co-creation, and to define their use case according to their relationship with technology.

## 7 Acknowlodgements

First and foremost, we would like to thank Andreas Breiter for his continuous support in our PhD and PostDoc projects. We are also grateful for the support of ZeMKI, in particular Andreas Hepp and Leif Kramp. In addition, we would like thank our colleagues at ifib, in particular Herbert Kubicek and Emese Stauke.